\documentclass[manuscript]{aastex}
\usepackage{natbib}
\slugcomment{}
\shorttitle{Dark Post-flare Loops}
\shortauthors{Song et al.}
\begin{document}
\title{Dark Post-flare Loops Observed by Solar Dynamics Observatory}

\author{Qiao Song$^{1, 2, 3}$, Jing-Song Wang$^{1}$, Xueshang Feng$^{2}$, and Xiaoxin Zhang$^{1}$}
\affil{1. Key Laboratory of Space Weather, National Center for Space Weather, China Meteorological Administration, Beijing 100081, China}
\affil{2. State Key Laboratory of Space Weather, National Space Science Center, Chinese Academy of Sciences, Beijing 100190, China}
\affil{3. Key Laboratory of Solar Activity, National Astronomical Observatories, Chinese Academy of Sciences, Beijing 100012, China}
\email{songq@cma.gov.cn}

\begin{abstract}
Solar post-flare loops (PFLs) are arcade-like loop systems that appear during the gradual phases of eruptive flares. The extreme ultraviolet (EUV) observations from the Atmospheric Imaging Assembly (AIA) on board the \emph{Solar Dynamics Observatory} (\emph{SDO}) allow us to investigate the fine structures in PFLs. In this work, we focus on studying the dark post-flare loops (DPFLs) during X-class flares, 
which are more evident in \emph{SDO}/AIA data than in previous EUV data. We identify and analyze the DPFLs observed by \emph{SDO} and find that: (1) the DPFLs of an X5.4 flare have an average lifetime of 10.0\,$\pm$\,5.5 minutes, an average width of 1022\,$\pm$\,339 km, and an average maximum length of 33\,$\pm$\,10 Mm, (2) blob-like falling features with a size close to the resolution of \emph{SDO}/AIA are identified in the DPFLs and they have an average velocity of 76\,$\pm$\,19 km~s$^{-1}$, and (3) the average widths of the DPFLs slightly increase with the characteristic temperatures in AIA 304, 171, 193, and 211 {\AA} channels. Our investigation shows that DPFLs are found in all of the 20 cases within this study,  which suggests that they are a common phenomenon in X-class flares and are probably produced by the same mechanism that creates coronal rain.
\end{abstract}


\section{Introduction}
    \label{S-Introduction}
An eruptive solar flare is a powerful and explosive phenomenon in the solar atmosphere which can release up to
10$^{32}$~ergs of energy and may lead to disastrous space weather. Observations show that a flare has three phases: a pre-flare phase, an impulsive phase, and a gradual phase \citep{2009soco.book.....G}. Post-flare loops (PFLs) are arcade-like, developed loop systems that appear in the gradual phases of eruptive flares.

PFLs were first found in H$\alpha$ images, where they appeared as bright loops above the solar limb \citep{1964ApJ...140..746B,1980sfsl.work..341M,1995SoPh..156..337S,1996SoPh..166...89W,1997SoPh..174..151V}. PFLs have also been observed in other wavelengths, such as ultraviolet, X-ray, and radio \citep{1975SoPh...45..377S,1977ApJ...214..891K}. The loop structures of hot X-ray and cool H$\alpha$ PFLs extend into the corona, with their footpoints identified as H$\alpha$ flare ribbons in the chromosphere \citep{1973SoPh...33..445R,1996mpsa.conf..211S}. For a particular wavelength, the loop structures of PFLs appear to rise and expand. For different wavelengths, hot X-ray PFLs occur earlier and higher, and cool H$\alpha$ PFLs show up later and lower, which indicates that the growth of PFLs is a superficial phenomenon of the cooling process \citep{1987SoPh..108..237S,2002AARv..10..313P}.

Two mechanisms generally contribute to the cooling process of PFLs: conduction and radiation. Cooling timescales for the two mechanisms can be calculated to determine which mechanism is dominating. The width and velocity of flare ribbons are used to estimate the cooling time due to conduction and radiation acting together \citep{1982SoPh...78..271S,2002AARv..10..313P}. If the parameters of the loops, such as length, temperature, and electron density, could be learned from the multi-wavelength observations or numerical modeling, the conductive and radiative cooling times could be calculated respectively \citep{1995SoPh..156..337S,2002SoPh..206..133S,2006ApJ...648..712H}. The observations and modeling \citep{1995ApJ...439.1034C, 2001SoPh..204...91A} have shown that the cooling process of PFLs has two phases: the conductive cooling, followed by the radiative coolings. 

Generally, twisted magnetic fields store enormous energy, and that energy will be released when a flare erupts \citep{1987ApJ...314..782G,2008ApJ...675.1637S,2013RAA....13..226S}. As a part of the flare process, PFLs are often explained as a result of magnetic reconnection in the standard flare model, i.e., the CSHKP model \citep{1964NASSP..50..451C,1966Natur.211..697S,1974SoPh...34..323H,1976SoPh...50...85K}. \citet{1996ApJ...459..330F} presented a scenario about the structure and the formation of the PFL system.
According to the scenario, the energetic particles from the magnetic reconnection of a flare can lead to the evaporation of the chromospheric material. Then, the chromospheric material expands into the corona along reconnected magnetic field lines and causes bright PFLs in high-temperature wavebands. As the hot plasma cools down to the chromospheric temperature, PFLs appear in H$\alpha$ images. Meanwhile, thermal equilibrium in the cooling loops is unstable, and the thermal instability \citep{1953ApJ...117..431P, 1965ApJ...142..531F} causes the formation of condensation downflows, which can be observed in the both legs of a PFL  \citep{1999SoPh..187..261S}.

Such downflow is a common phenomenon on the Sun. In non-flare loops of active regions (ARs), thermal instability leads to catastrophic cooling and the formation of cool and dense plasma blobs \citep{2001SoPh..198..325S}. These blobs, which are called coronal rain, are often observed and increasingly studied \citep{1970PASJ...22..405K,1972SoPh...25..413L,1976ApJ...210..575F,1978ApJ...223.1046F,2007A&A...475L..25O,2009ApJ...694.1256T,2009ApJ...695..642U,2010ApJ...716..154A,2011A&A...532A..96K,2012SoPh..280..457A,2015ApJ...806...81A}.  Generally, downflows in PLFs appear as dark loop-like features in extreme ultraviolet (EUV) images,  which are caused by the absorption of hydrogen and helium in the cooling process \citep{2005ApJ...622..714A}. Thus, they are called dark post-flare loops (DPFLs) in this work. 

Although DPFLs have been observed for years, little attention has been paid to their fine structures in EUV wavelengths and their relationship to coronal rain \citep{2014ApJ...797...36S,2015ApJ...806...81A}. Now, the unprecedented observations from the Atmospheric Imaging Assembly (AIA; \citealt{2012SoPh..275...17L}) on board the \emph{Solar Dynamics Observatory } (\emph{SDO}; \citealt{2012SoPh..275....3P}) allow us to investigate the fine structures of DPFLs in multiple EUV wavelengths. In this work, based on \emph{SDO}/AIA data, we study DPFLs and analyze their properties with the multiband EUV observations of the X-class flares in Solar Cycle 24. After presenting our data in Section~\ref{Data}, we describe the observational characteristics of DPFLs and calculate the cooling time in Section~\ref{Results}. In Section~\ref{C&D}, we summarize this work and give a brief discussion.

\section{Data}
    \label{Data}
In this work, we investigated 20 GOES X-class flares from 2011 February to 2014 February. The flare list in Table~\ref{tab:1} is taken from \citet{2014ApJ...782L..27Z} with permission. It contains all 19 GOES X-class flares between 2011 February and 2013 August and lists their start, peak, and end times, GOES classes, and the corresponding AR characteristics. Two X-class flares happened in AR 11429 on 2012 March 7 and an X4.9 flare occurred in AR 11990 on 2014 February 25. We chose AR 11429 as a target to investigate the evolution of DPFLs on the solar disk in detail, and another target, AR 11990, was chosen to study DPFLs on the limb.

Without atmospheric disturbance, \emph{SDO} provides high-quality images
of the Sun. With the high spatial resolution ($\sim1''$) and high cadence (12 s) full-disk data observed
by \emph{SDO}/AIA, we can study the coronal dynamics in unprecedented detail. \emph{SDO}/AIA provides
seven EUV wavelengths (304, 171, 193, 211, 335, 94, and 131 {\AA}) which cover a characteristic temperature
range from $\sim5\times10^{4}$ to $\sim2\times10^{7}$ K. 
In this work, we studied the events in all the seven EUV channels, especially in 171 {\AA} for its best quality in
observations of DPFLs. We also used simultaneous longitudinal magnetograms from \emph{SDO}/HMI to find the magnetic structures related to DPFLs. The standard \emph{SDO} routines (aia\_prep.pro and hmi\_prep.pro) in the Solar SoftWare (SSW) were used to convert level 1 data to level 1.5. The time-series data on the disk used here were corrected for the solar rotation.

\section{Results}
    \label{Results}

\subsection{DPFLs on disk}
    \label{dod}
\begin{figure}[htbp]\centering
\includegraphics[width=0.8\textwidth,clip]{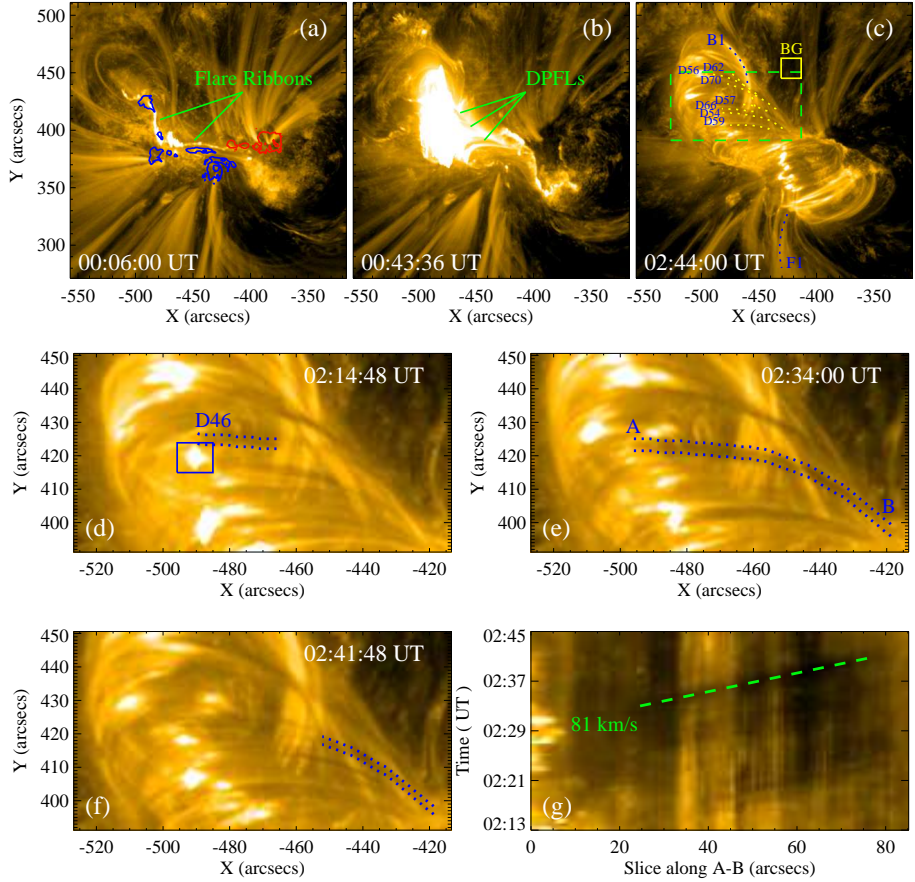}
\caption{(a)-(c) Overview of dark post-flare loops (DPFLs) of NOAA active region 11429 in the 171 {\AA}
channel on \emph{SDO}/AIA. The red (blue) contours in panel (a) are the positive (negative) longitudinal
magnetic fields with the absolute levels of 1000 and 1500 G. The yellow dotted curves in panel (c) present
identified DPFLs (D54, D56, D57, D59, D62, D66, and D70). B1 is a bright post-flare loop and F1 is a fan-like loop of
the active region. The BG box displays a background region. The green dashed box shows the field of view of panels
(d)-(f), which display the evolution of a typical DPFL (D46). Panel (g)
is a space-time plot which is made by the time slice along the middle line of A-B shown in the panel (e). The
slope of the dashed line means an apparent velocity of $\sim$ 81 km~$s^{-1}$ of a blob-like feature in the DPFL (a color version is
available online).}\label{fig:1}
\end{figure}

On 2012 March 7, two X-class flares happened in AR 11429. The first one (X5.4) took
place at the northeastern part of the AR at 00:02 UT and the second one (X1.3) erupted at 01:05 UT in southwest.
The X5.4 flare was an eruptive flare with a typical two-ribbon structure. Figure~\ref{fig:1}a shows
the bright flare ribbons at 00:06 UT in the 171 {\AA} channel on \emph{SDO}/AIA with the contours of longitudinal magnetic
fields ($\pm$1000 and $\pm$1500 G) in blue and red colors. The east and the west flare ribbons were located in the negative magnetic polarity (blue) and the positive magnetic polarity (red), respectively. The flare ribbons behaved as bright patches because of saturation at the flare peak time 00:24 UT. About 19 minutes later, the DPFLs, the dark and loop-like features, gradually became clear among bright patches of the flare (see Figure~\ref{fig:1}b). Seven identified DPFLs (D54, D56, D57, D59, D62, D66, and D70) at 02:44 UT are indicated by yellow dotted curves in Figure~\ref{fig:1}c, and multiple DPFLs were frequently observed in different loops during the evolution of the PFL system. After a four-hour evolution and expansion, the PFLs became
dispersed and finally disappeared within half a day (see the associated movie).

We tracked each identifiable DPFL of the X5.4 flare from $\sim$00:50 UT, when the first
one appeared, to $\sim$04:50 UT, when the last one disappeared, in the 171 {\AA} channel. A total of 120
DPFLs were identified during these four hours. We measured the EUV brightness of the DPFLs almost simultaneously in the 304, 171, 193, and 211 {\AA}  images. As shown in Figure~\ref{fig:1}c, we chose a small area, the background region (BG), which is without any significant coronal loops. Compared to bright loops, a DPFL is a dark structure, but it is still brighter than the background. For example, the brightness of a typical DPFL (D46) was only about one-fifth of that of a bright PFL (B1) in each of the four channels. However, D46 was eight times brighter than the background region (BG) and had a similar brightness as that of a fan-like loop (F1) in each of the four channels.

The evolution of D46 is shown in Figures~\ref{fig:1} (d-f), with the same field of view outlined by the green dashed box in Figure~\ref{fig:1}c. D46 formed after a bright knot appeared near the top of the PFLs (see the small box in Figure~\ref{fig:1}d). As dark downflows fell down, the shape of D46 changed dramatically and stretched to the western footpoint of the PFL. At 02:34 UT, the projected length of D46 was about 44 Mm. It shortened from the top of the loop and eventually disappeared about 7 minutes later (see Figure~\ref{fig:1}f). 

It is noteworthy that the DPFLs contained tiny structures. A space-time plot was made through a slice (A-B) along the centerline of D46 (see Figures~\ref{fig:1}e and g). The dark lines in Figure~\ref{fig:1}g show some blob-like falling features which were even darker than the other parts of D46, and  the size of the blob-like features were close to one pixel ($\sim435$ km) in the image. The apparent velocity (projected in the plane of the sky) of the blob-like features was 81 km~s$^{-1}$ as shown by the dashed line in Figure~\ref{fig:1}g. These dark parallel lines below the dashed line show earlier dark blob-like features in the same DPFL with similar velocities. We found that the blob-like features appeared in all the DPFLs and usually a DPFL contained more than one blob-like feature. The apparent velocities of the blob-like features were measured as an indicator of the velocities of DPFLs.

\begin{figure}[htbp]\centering
\includegraphics[width=1\textwidth,clip]{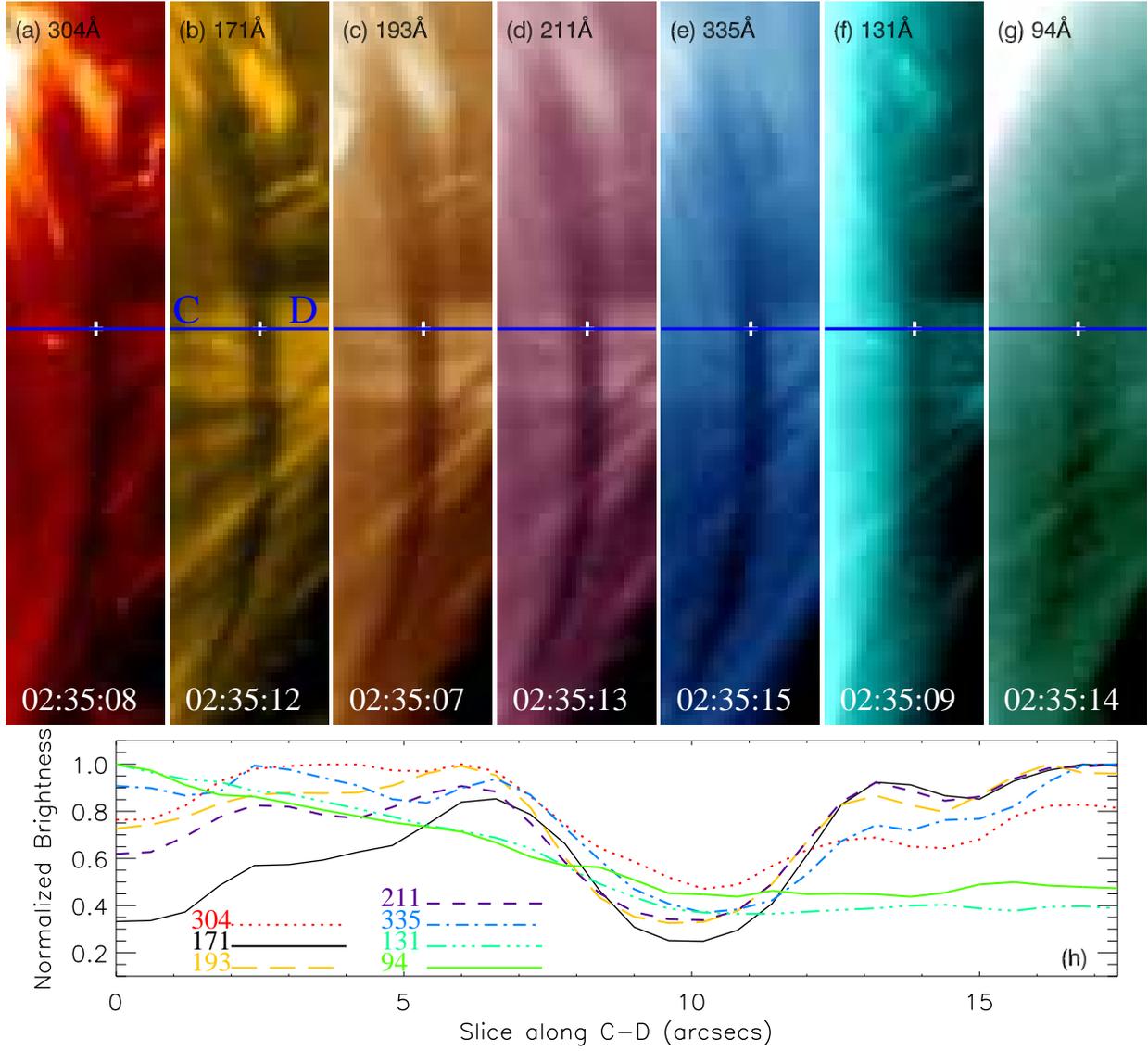}
\caption{(a)-(g): Multiband observations of a DPFL (D46) at 02:35 UT on 2012 March 7. (h) The brightness variation along slice C-D in the seven AIA EUV channels.}\label{fig:2}
\end{figure}

Most DPFLs during the X5.4 flare on 2012 March 7 appeared in the 304, 171, 193, and 211 {\AA} channels, while parts of DPFLs could be observed in all the seven EUV channels of AIA. Figures~\ref{fig:2} (a-g) show D46 as a multiband example at~02:35 UT. D46, especially the upper half of the loop, was less clear in the three high temperatures channels, 335, 131, and 94 {\AA},  than the other four cooler channels. The difference between the upper and lower halves of the loop in the multiple EUV channels may reflect the longitudinal differences of the temperature and density in the loop. It is particularly difficult to detect the upper part of the DPFL in the 94 {\AA} channel (log T = 6.8), which may indicate that it had been cooled below the range of temperature response in the AIA 94 {\AA} channel. Figure~\ref{fig:2}h shows the normalized EUV brightness variation along the slice C-D in the seven channels. In the 171, 193, and 211 {\AA} channels, the brightness of D46 was less than half of that of its surroundings, which made D46 appear clearly and become more easily identifiable in these channels than in the other channels. 

Table~\ref{tab:3} gives the average widths of the DPFLs in six AIA EUV channels, 304, 171, 193, 211, 335, and 131 {\AA}, which have roughly rising characteristic temperatures. To compare with the previous results, the first six rows of widths were calculated through a semi-automatic procedure, which consists of tracking a loop manually, measuring the brightness variation in the cross-cutting of the loop, extracting the absorption curve from the background of the brightness variation, and counting the full width half maximum (FWHM) of the curve automatically. The standard deviation of the widths and the number of samples are given in the third and fourth columns, respectively. The curves that under the effect of saturation or having small depths were excluded from the sample. The sample number of the 193 {\AA} channel was significantly less than those of the 171 and 211 {\AA} channels because the PFLs in the 193  {\AA} channel were saturated longer than the PFLs in the 171 and 211 {\AA} channels. Roughly, fewer sample numbers of a channel means that it is harder to find a specific DPFL in the channel. We found that the average widths are increasing with the characteristic temperatures in the first four AIA channels, 304, 171, 193, and 211  {\AA}, but the difference of widths among the different channels was within the range of the standard deviations. This rising trend was discontinued in the remaining two channels. The width of the AIA 335 {\AA} channel was slightly less than that of the 211 {\AA} channel, and the width in the 131 {\AA} channel was less than those of the other five channels. The 131 {\AA} channel contains both high and low temperature components, which may explain its small width, large deviation, and small sample number. The difference in images, widths, and sample numbers may suggest that the DPFLs have a transverse multi-temperature structures within a specific temperature range. The width in the last row, 1022\,$\pm$\,339 km, was measured manually as a reference, which reveals that the automatic widths were systematically larger than the manual widths. The sample numbers from the automatic procedure were also less than that of the manual procedure, which show that the manual procedure is important when deal with the complex situation of some DPFLs.

\begin{figure}[htbp]\centering
\includegraphics[width=1\textwidth,clip]{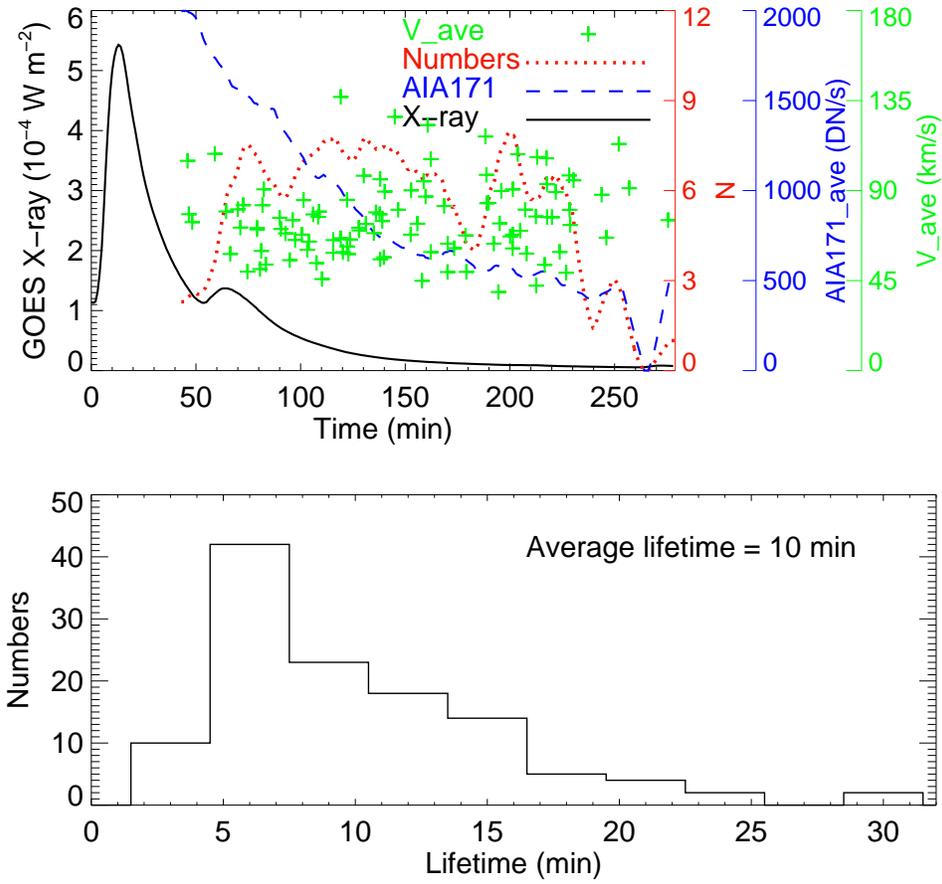}
\caption{Upper panel: evolution of brightness (AIA171\_ave), numbers of the DPFLs, and GOES soft X-ray flux. The average downflow velocities (V\_ave) of the DPFLs are presented by the green plus signs with their x-coordinates according to the middle time of the lifetime of the DPFLs. Lower panel: lifetime distribution of the
DPFLs.}\label{fig:3}
\end{figure}

In addition to the width, we made statistical analyses for other characteristics of the 120 identified DPFLs in the X5.4 flare in the 171 {\AA}  channel. The 120 DPFLs had an average maximum length of  33\,$\pm$\,10 Mm and an average downflow velocity of 76\,$\pm$\,19 km~s$^{-1}$, respectively. As shown in the upper panel of Figure~\ref{fig:3}, the average brightness of the DPFLs in the 171 {\AA} channel decreased coincidentally with the
GOES soft X-ray flux, while the average downflow velocities and numbers of the DPFLs did not change monotonously with time. The number of the DPFLs remained relatively steady and then decreased significantly $\sim$240 minutes after the flare erupted, which indicates that multiple DPFLs are often observed at the same time. The histogram in the lower panel of Figure~\ref{fig:3} illustrates that one-third of the DPFLs had a typical lifetime of about 6 minutes and the average lifetime was 10.0\,$\pm$\,5.5 minutes.

\subsection{DPFLs on limb}
    \label{dol}
\begin{figure}[htbp]\centering
\includegraphics[width=0.9\textwidth,clip]{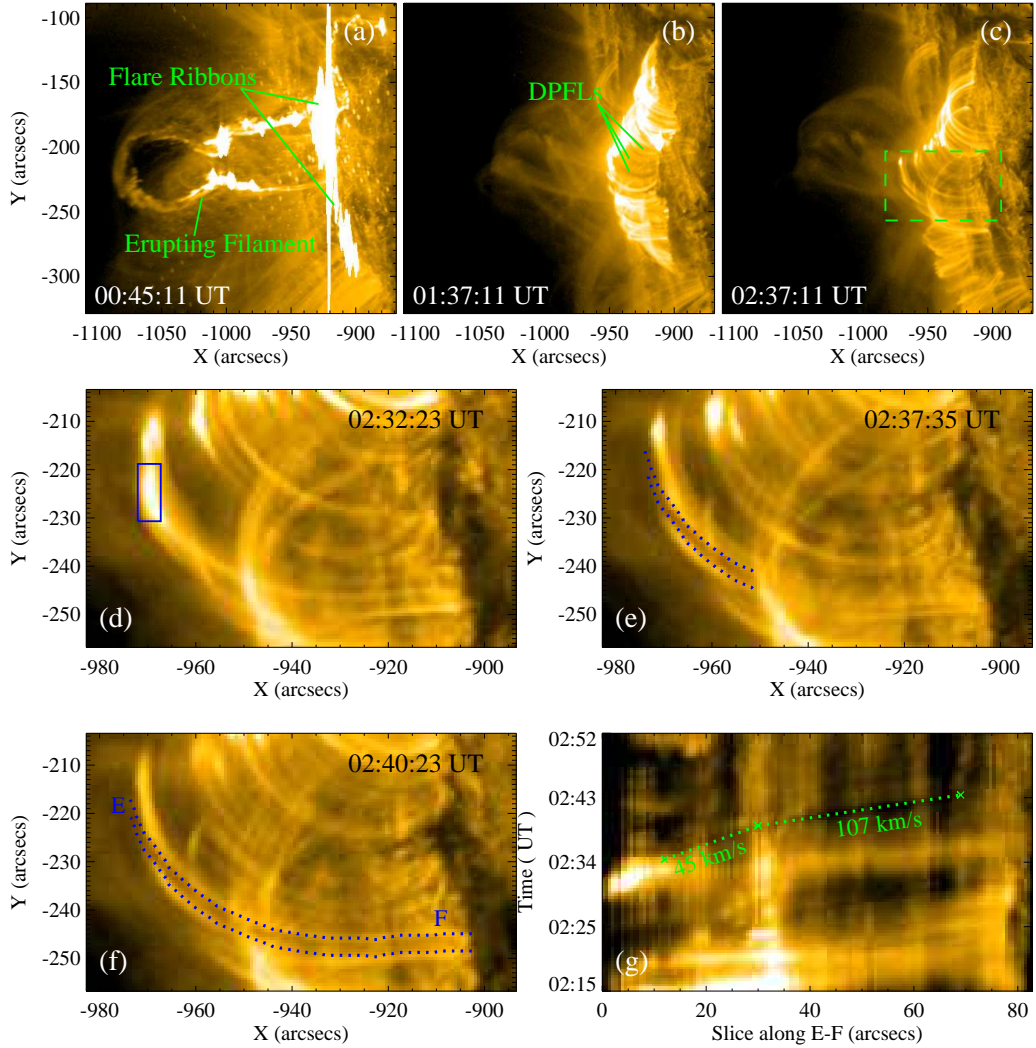}
\caption{(a)-(c) Overview of DPFLs of AR 11990 on the east limb on 2014 February 25 in the AIA 171 {\AA} channel. The green dash box in panel (c) shows the field of view of panels (d)-(f), which display the evolution of a typical DPFL. The panel (g) is a space-time plot which is made by the time slice along the middle line of E-F shown in the panel (f). The slopes of the dashed lines mean that velocity of a blob-like feature in the DPFL increases from 45 to 107 km~$s^{-1}$.}\label{fig:4}
\end{figure}

Solar limb events give us another viewpoint to check the properties of DPFLs. On the east solar limb, an X4.9 flare erupted in AR 11990 at 00:39 UT on 2014 February 25. The X4.9 flare was a typical eruptive event with peak time at 00:49 and end time at 01:03 UT. Figures~\ref{fig:4} (a-c) show the evolution of the flare event in the AIA 171 {\AA} channel. After the flare erupted, a filament rose slowly at 00:42 UT, and then the overlying loops opened in the field of view after the filament's quick eruption at 00:45 UT. At the same time, flare ribbons appeared and became saturated, owing to high energy in the AIA 171 {\AA} channel. When the saturation started to disappear, the first DPFL appeared around 01:10 UT. Figure~\ref{fig:4}b shows that some DPFLs can be clearly identified in the PFL system at 01:37 UT. Most of DPFLs disappeared before 03:10 UT, and the whole PFL system dispersed at $\sim$04:00 UT.

Figures~\ref{fig:4} (d-f) display the evolution of a typical DPFL on the solar limb, with the same field of
view outlined by the green dashed box in Figure~\ref{fig:4}c. Similar to the on-disk case, the DPFL appeared after a bright knot
near the top of the PFLs (see the small box in Figure~\ref{fig:4}d). Then the DPFL stretched to the western footpoint as downflows fell down (see Figure~\ref{fig:4}f). About 10 minutes after the DPFL appeared, it shortened and eventually disappeared. A slice (E-F) along the centerline of the DPFL is used to make a space-time plot (see Figure~\ref{fig:4}g). The dotted lines in Figure~\ref{fig:4}g show that a blob-like feature in the DPFL was accelerated from $\sim$45 km~s$^{-1}$ to $\sim$107 km~s$^{-1}$. Some dark lines parallel to the dotted lines show the other earlier or later blob-like features in the same DPFL and indicate similar velocities.

\begin{figure}[htbp]\centering
\includegraphics[width=0.9\textwidth,clip]{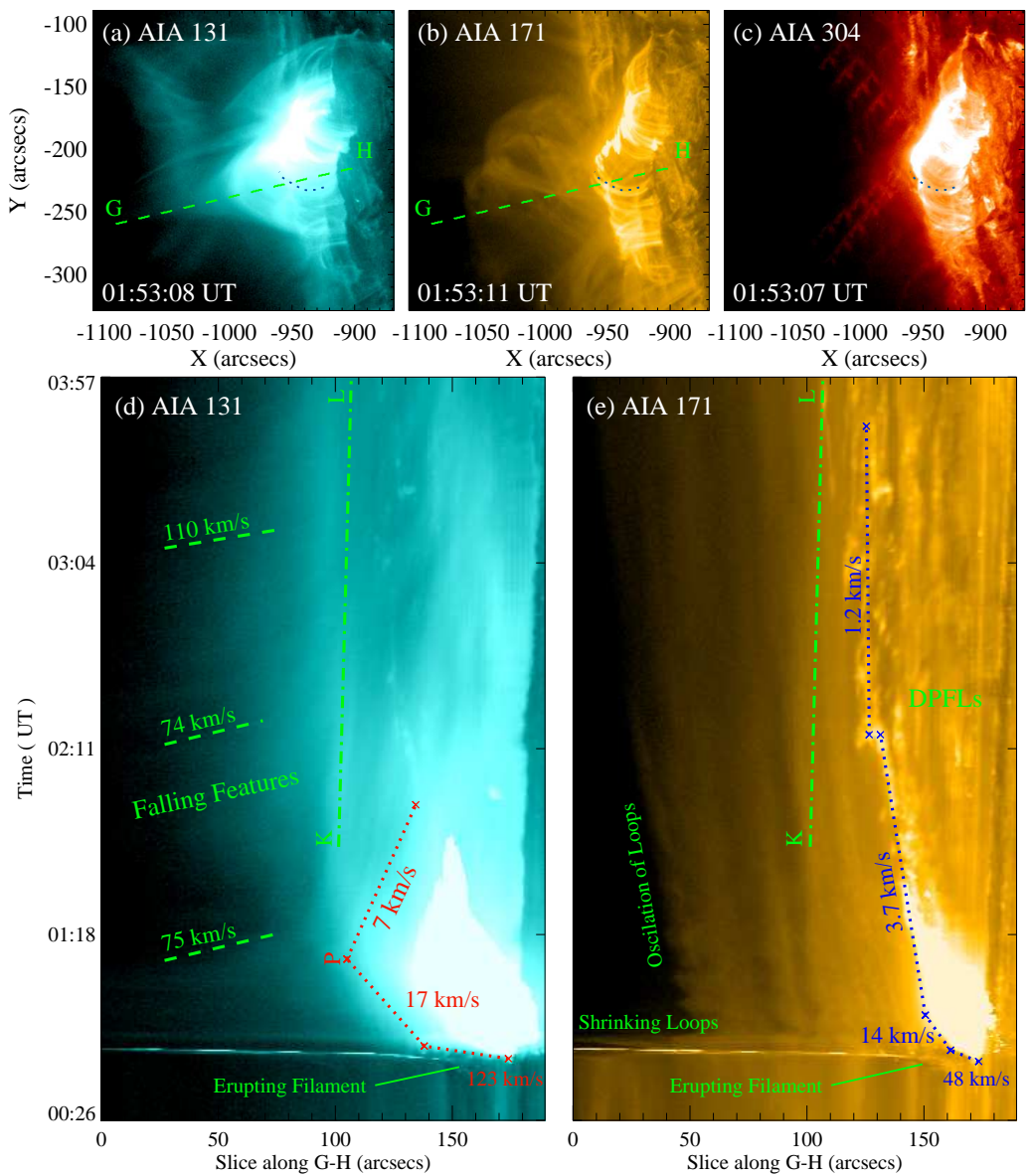}
\caption{(a)-(c) \emph{SDO}/AIA multi-wavelength images of DPFLs in AR 11990 on 2014 February 25. The blue dotted curves in each panel present the location of a DPFL. The dashed lines in panels (a)-(b) indicate the position of a slice (G-H) for panels (d)-(e). (d)-(e) Space-time plots made by the slice G-H in the AIA 131 and 171 {\AA} channels, respectively.}\label{fig:5}
\end{figure}

Figure~\ref{fig:5} displays the various appearances of DPFLs in three different AIA channels (131, 171, and 304 {\AA}) and shows the loop structures at different heights from the side. As shown in Figure~\ref{fig:5}a, bright PFLs were expanded above the solar limb, and a cusp was located above the PFLs in the AIA 131 {\AA} channel. Some high-speed falling features, which were also above the PFLs, were observed around the cusp in high corona. The DPFLs in the 171 {\AA} channel were clearer than those in other channels (see the dotted line in Figure~\ref{fig:5}b). The DPFLs and the high loops were weak in the cool 304 image (log T = 4.7) from the side view in Figure~\ref{fig:5}c.

We studied the evolution of loop structures at different heights using a slice (G-H) in the 131 and 171 {\AA} channel. The slice is throughout the cusp and approximately perpendicular to the top of the PFLs (see the dashed lines in Figures~\ref{fig:5}a and b). Figures~\ref{fig:5} (d-e) show space-time plots which are made by the slice G-H in the 131 and 171 {\AA} channels, respectively. Each space-time plot is divided into two parts that respectively represent the high and low corona, by the dashed and dotted line (K-L).

In these two channels, the evolution of loop structures is distinct at different heights. In the high corona, loops shrank and oscillated along the slice, and the amplitude of the oscillation decreased with time in the 171 {\AA} channel. Some high-speed falling features with velocities of $\sim$100~km~s$^{-1}$ were observed in the 131 {\AA} channel. The effect of solar rotation on the calculation of the velocities was negated by the track of loop footpoints. In the low corona, the PFLs in the 131 {\AA} channel rose rapidly with an apparent velocity ($V_{a}$) of 123 km~s$^{-1}$, then slowed to 17 km~s$^{-1}$ before the point P, and finally turned to a speed of 7 km~s$^{-1}$ with the opposite direction. On the other hand, $V_{a}$ of the PFLs in the AIA 171 {\AA} channel could be divided into at least four stages. $V_{a}$ was about 48~km~s$^{-1}$ in the first stage, decreased to 14~km~s$^{-1}$ in the second stage, continuously decreased to an average $V_{a}$ of 3.7~km~s$^{-1}$ in the third stage, and finally became to 1.2~km~s$^{-1}$ in the last stage. DPFLs appeared in the third and fourth stage, which belong to the decay phase (the latter part of gradual phase) of the flare. So $V_{a}$ was 3.7--1.2~km~s$^{-1}$ during the period of the DPFLs, and these velocities are used to calculate the cooling time in Section~\ref{Cooling process}.

Besides the distinctions, we also found some connections between different heights and channels. The erupting filament appeared in both the 131 and 171 {\AA} channels. Furthermore, the evolution of the DPFLs is related to the movement of the PFL system. The DPFLs in the 171 {\AA} channel started to appear when the $V_{a}$ in the 131 {\AA} channel turned to the opposite direction at $\sim$01:10 UT (see the point P in Figure~\ref{fig:5}d). Most DPFLs disappeared around 03:10 UT, when $V_{a}$ was approximately zero and the top of the PFLs was close to the dividing line K-L. On the other side of the dividing line, falling features in the 131 {\AA} channel decelerated as they get closer to the dividing line. After the effect of solar rotation was removed, we found that the height of the dividing line was not change with time.

\subsection{Cooling process of DPFLs}
\label{Cooling process}
As mentioned above, DPFLs are derived from the cooling process of a PFL system which mainly includes two mechanisms, conduction and radiation. We can determine which mechanism predominates by respectively calculating the conductive and radiative cooling time \citep{2002SoPh..206..133S}.
The conductive cooling timescale ($\tau_{cond}$) of the loops is calculated by
\begin{equation}\label{eq:2}
\tau_{cond} = \frac{3n_{e}k_{B}L^2}{\kappa_{0}T^{5/2}},
\end{equation}
where $k_{B}=1.38\times10^{-16}$~erg~K$^{-1}$ is the Boltzman constant and $\kappa_{0}=9.2\times10^{-7}$~erg~s$^{-1}$~cm$^{-1}$~K$^{-7/2}$ is the thermal Spitzer conductivity coefficient \citep{1962pfig.book.....S}. $T$, $n_{e}$, and $L$ are the temperature, electron density, and half-length of loops, respectively. Table~\ref{tab:2} gives some previous results of the temperature, electron density, half-length, cooling time, and ascending velocity of PFLs. Since values of physical parameters ($T$, $n_{e}$, and $L$) depend on observational wavelengths and change greatly during a flare process \citep{2003SoPh..215..127K}, here we select $T =1\times10^{6}$~K and $n_{e}=1\times10^{10}$~cm$^{-3}$ based on the references in Table~\ref{tab:2} and the characteristic temperature of the AIA 171 {\AA} channel \citep{2012SoPh..275...17L}. We choose $L=2.5\times10^{9}$~cm as a lower limit value of $L$ to get a lower limit of $\tau_{cond}$. By substituting the parameters to Equation~\ref{eq:2}, we obtain $\tau_{cond}=2.8\times10^4$~s, which is still longer than the duration of the flare process.

On the other hand, under the hypothesis of full ionization and with the piece-wide power law approximation from
\citet{1978ApJ...220..643R}, the radiative cooling timescale ($\tau_{rad}$) is determined by
\begin{equation}\label{eq:3}
\tau_{rad} = \frac{3k_{B}T}{n_{e}\Lambda_{0}},
\end{equation}
where $\Lambda_{0}=1.2\times10^{-22}$~erg~s$^{-1}$~cm$^{3}$ is the radiative loss rate at $1\times10^{6}$~K \citep{2004psci.book.....A}. The result is $\tau_{rad}=345$~s. In Section~\ref{dod}, we found that the DPFLs have a typical lifetime of 6 minutes, which is close to $\tau_{rad}$.

Moreover, it was found that the cooling time can be estimated by using the velocity and the width of the flare ribbons \citep{1982SoPh...78..271S,2002AARv..10..313P}. The PFLs are observed expanding in a particular wavelength during the cooling process. \citet{1995SoPh..156..337S} suggested that the cooling time of loops is approximately equal to the characteristic growing time of PFLs. So we can estimate the cooling timescale in a particular wavelength by using the movement of PFLs. The cooling timescale ($\tau_{cr}$) is estimated by
\begin{equation}\label{eq:1}
\tau_{cr} \approx W/V_{a},
\end{equation}
where $W$ is the width of the loop and $V_{a}$ is the ascending velocity of the PFLs. In this work, $W=1022$~km and $V_{a}=3.7-1.2$~km~s$^{-1}$, and we obtain $\tau_{cr}\approx276-852$~s, which reflects a combination of the two cooling mechanisms in the AIA 171 {\AA} channel. The results show that lifetime and $\tau_{cr}$ are comparable with $\tau_{rad}$ but are much less than $\tau_{cond}$. Therefore, the cooling process of DPFLs is dominated by the radiation and the conductive cooling for DPFLs can be neglected during the decaying phase of an X-class flare. 

\subsection{DPFLs of X-class flares}
    \label{doxf}
\begin{figure}[htbp]\centering
\includegraphics[width=1\textwidth,clip]{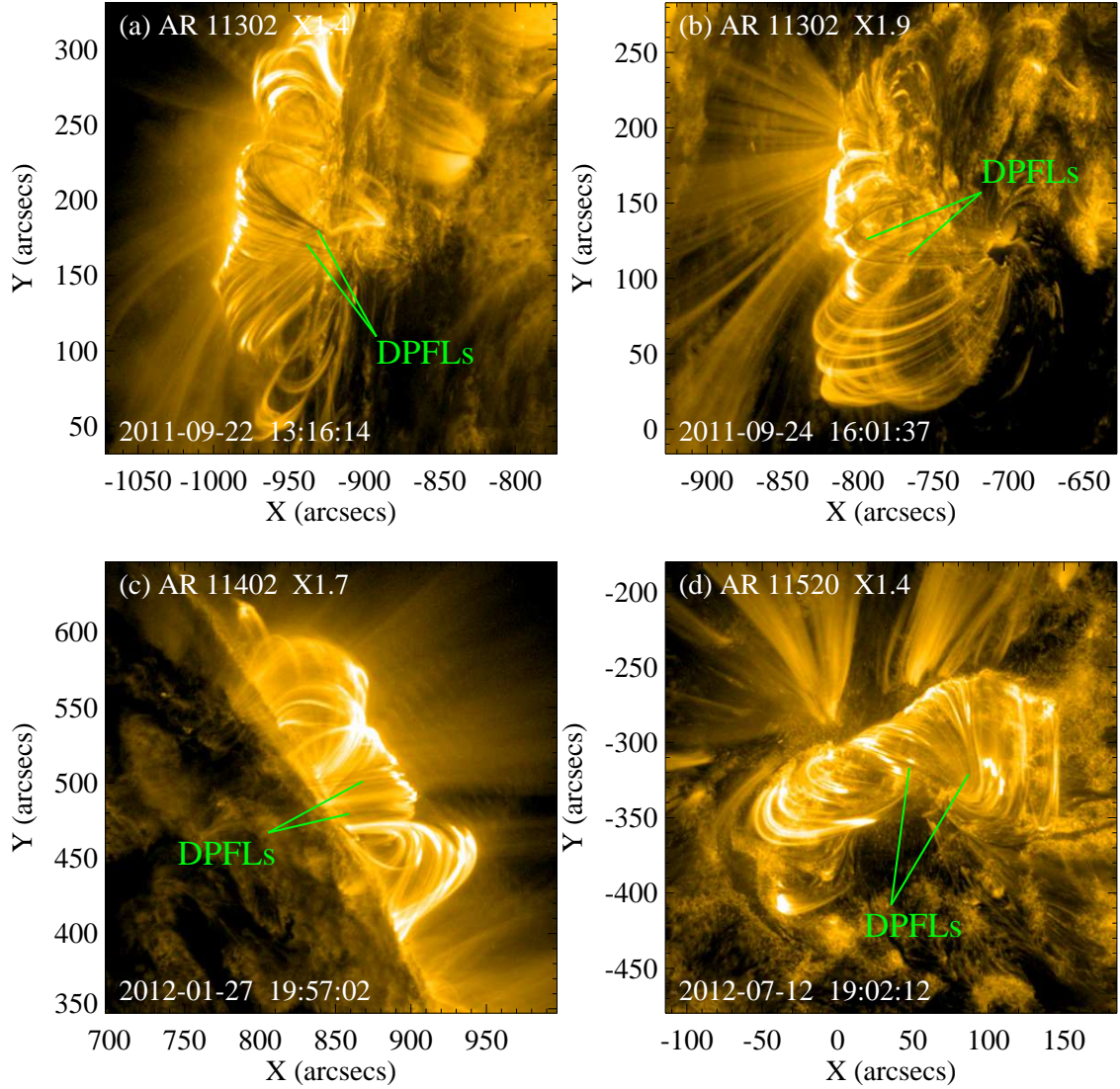}
\caption{DPFLs of four X-class flares observed in the 171 {\AA} channel of \emph{SDO}/AIA. (a) DPFLs of an X1.4 flare that happened in AR 11302 on 2011 September 22. (b) DPFLs of an X1.9 flare in the same AR in two days later. (c)-(d) DPFLs of an X1.7 and an X1.4 flares which erupted in AR 11402 and AR 11520 on 2012 January 27 and 2012 July 12, respectively.}\label{fig:6}
\end{figure}

Twenty GOES X-class flares between 2011 February and 2014 February were investigated and had similar features for DPFLs. The DPFLs can be identified in all the 20 X-class flares and their appearance and disappearance time are listed in Table~\ref{tab:1}. Most of them appeared after the end of the GOES flares, and then they developed during an average period of 4 hr, and finally disappeared. Figure~\ref{fig:6} gives snapshots of the DPFLs during four flares in the 171 {\AA} channel of AIA. As shown in Figure~\ref{fig:6}a, the PFLs of an eruptive X1.4 flare that occurred in AR 11302 at 10:29 UT expanded above the east limb of the Sun on 2011 September 22. The first DPFL appeared about one hour after the X1.4 flare erupted, and the last one disappeared about nine hours after the flare erupted. Figure~\ref{fig:6}b shows the DPFLs of an X1.9 flare in the same AR at 16:01 UT, 2011 September 24. The X1.9 flare erupted at 09:21 UT with complicated PFLs. Figure~\ref{fig:6}c shows that the DPFLs of an X1.7 flare at 19:57 UT on 2012 January 27. This flare began at 17:37 UT in AR 11402 above the west limb of the Sun. The fourth example (Figure~\ref{fig:6}d) is an X1.4 flare that erupted at 15:37 on 2012 July 12 in AR 11520 near
the center of the solar disk. The DPFLs emerged at $\sim$17:30 UT and disappeared at $\sim$23:29 UT. 

\section{Discussions and Conclusions}
    \label{C&D}

Using the EUV observations from \emph{SDO}/AIA, we studied DPFLs in eruptive flares and found blob-like features in DPFLs. In this work, the properties of DPFLs were investigated by analyzing the EUV data with high spatial and temporal resolutions. The main results are summarized as follows:

\begin{itemize}

   \item[1.] DPFLs were found in 20 X-class flares between 2011 February and 2014 February. Most of the DPFLs appeared after the end of the GOES flares, and lasted for a period of about 4 hr.
   
   \item[2.] The DPFLs were most distinct in the \emph{SDO}/AIA EUV 171 {\AA} channel. Most of the DPFLs were noticeable in the images of 304, 171, 193, and 211 {\AA}, while some of them could be found in all EUV wavelengths of \emph{SDO}/AIA.
   
   \item[3.] The EUV brightness of the DPFLs was actually brighter than that of the background region, with an evolution similar to that of GOES X-ray flux.

  \item[4.] The DPFLs of the X5.4 flare had an average lifetime of 10.0\,$\pm$\,5.5 minutes, an average width of 1022\,$\pm$\,339 km, and an average maximum length of 33\,$\pm$\,10 Mm.
  
  \item[5.] The DPFLs contained multiple blob-like features with a size close to the resolution of SDO/AIA. The blob-like features in the DPFLs of the X5.4 flare had an average velocity of 76\,$\pm$\,19 km~s$^{-1}$. 
 
   \item[6.] The cooling time of the X4.9 flare was 276--852 s during its decaying phase, and the cooling process was dominated by the radiation.  
   
   \item[7.] The average widths of the DPFLs slightly increase with the characteristic temperatures in the AIA 304, 171, 193, and 211 {\AA} channels.
   
   \item[8.] The comparison among multiple EUV channels reflects the presence of longitudinal and transverse temperature structures in the DPFLs.

\end{itemize}
In this work, we observed DPFLs in all the sample flares, which suggests DPFLs are a common phenomenon in the decay phase of X-class flare events. The results confirmed that the DPFLs originate from the cooling process due to the thermal instability in PFLs. The blob-like features in DPFLs behaved like coronal rain in non-flare AR loops \citep{1976ApJ...210..575F,2011A&A...532A..96K,2012SoPh..280..457A,2014ApJ...797...36S,2015ApJ...806...81A}. These blob-like features are most likely produced by the same mechanism that creates coronal rain, but with a different amount of mass from the chromospheric evaporation. The relationship between the widths of DPFLs and the characteristic temperatures of the EUV channels is consistent with the pervious results, and it suggests the existence of multi-temperature structures and a transition of temperature from the chromosphere to the corona \citep{2015ApJ...806...81A}. Considering their locations and velocities, the high-speed falling features above the PFLs in the AIA 131 images are probably supra-arcade downflows \citep{1999ApJ...519L..93M,2004ApJ...616.1224S,2011ApJ...730...98S,2012ApJ...747L..40S,2013MNRAS.434.1309L}. Supra-arcade downflows and DPFLs have an apparent correlation in time and location, but it is unclear if they are intrinsically related due to the magnetic reconnection. Future investigations are needed to answer these questions.

\acknowledgments
The authors gratefully thank the anonymous referee for insightful comments that helped us make a significant improvement of the manuscript. We also sincerely thank Prof. M. Zhang for fruitful advice and discussions that improved this work. We appreciate Prof. J. Zhang, Dr. Y.Z. Zhang, and Dr. G.P. Zhou for inspiring the original idea and giving much useful advice. We would like to thank Prof. P. F. Chen and Prof. M.J. Aschwanden for helpful discussions. We acknowledge Dr. S.H. Yang and Dr. K. Zhang for the careful reading and constructive advice.  The data have been used by courtesy of NASA/\emph{SDO} and the AIA science team. \emph{SDO} is a mission for NASA's Living With a Star program. The work is supported by the the National High-tech R\&D Program of China (2012AA121000), the National Basic Research Program of China (2011CB811403), and the National Natural Science Foundation of China (41404136, 40931056, 11125314, 11221063 and 11373004).
\begin{table}[htbp]
\begin{center}
\caption{List of 20 X-class flares in Solar Cycle 24 \citep{2014ApJ...782L..27Z} and DPFL time of the flares.} \label{tab:1}
\begin{tabular}{rccccccccccc}
\tableline\tableline
Date~~~~ & \multicolumn{3}{c}{Flare Time} & Flare & \multicolumn{2}{c}{Active Region} &~&\multicolumn{2}{c}{DPFL Time\tablenotemark{b}}\\
\cline{2-4}\cline{6-7}\cline{9-10}
(UT)~~~~~& Start & Peak & End & Class & Number & Location\tablenotemark{a}&~& Start & End \\
\tableline
   2011 Feb 15  & 01:44 & 01:56 & 02:06 & X2.2 & 11158 & S21W14 &~& 02:38 & 06:06 \\
 --     Mar ~9 & 23:13 & 23:23 & 23:29 & X1.5 & 11166 & N11W01 &~& 00:00 & 01:45 \\
 --     Aug ~9 & 07:48 & 08:05 & 08:08 & X6.9 & 11263 & N18W68 &~& 08:27 & 10:28 \\
 --     Sep ~6 & 22:12 & 22:20 & 22:24 & X2.1 & 11283 & N14W04 &~& 22:59 & 23:57 \\
 --     Sep ~7 & 22:32 & 22:38 & 22:44 & X1.8 & 11283 & N14W18 &~& 23:29 & 01:59 \\
 --     Sep 22 & 10:29 & 11:01 & 11:44 & X1.4 & 11302 & N11E74 &~& 11:15 & 19:58 \\
 --     Sep 24 & 09:21 & 09:40 & 09:48 & X1.9 & 11302 & N13E59 &~& 10:14 & 11:10 \\
 --     Nov ~3 & 20:16 & 20:27 & 20:32 & X1.9 & 11339 & N18E63 &~& 21:22 & 00:41 \\
  2012 Jan 27 & 17:37 & 18:37 & 18:56 & X1.7 & 11402 & N29W72 &~& 19:16 & 23:57 \\
 --    Mar ~5 & 02:30 & 04:09 & 04:43 & X1.1 & 11429 & N18E55 &~& 04:29 & 09:57 \\
 --    Mar ~7 & 00:02 & 00:24 & 00:40 & X5.4 & 11429 & N17E29 &~& 00:50 & 04:50 \\
 --    Mar ~7 & 01:05 & 01:14 & 01:23 & X1.3 & 11429 & N17E29 &~& 01:26 & 09:30 \\
 --    Jul ~6 & 23:01 & 23:08 & 23:14 & X1.1 & 11515 & S17W36 &~& 23:45 & 01:15 \\
 --    Jul 12 & 15:37 & 16:49 & 17:30 & X1.4 & 11520 & S17E06 &~& 17:30 & 23:29 \\
 --    Oct 23 & 03:13 & 03:17 & 03:21 & X1.8 & 11598 & S10E56 &~& 04:29 & 06:30 \\
  2013 May 13 & 01:53 & 02:17 & 02:23 & X1.7 & 11748 & N12E94 &~& 02:46 & 07:45 \\
 --    May 13 & 15:48 & 16:05 & 16:16 & X2.8 & 11748 & N12E82 &~& 16:45 & 21:45 \\
 --    May 13 & 23:59 & 01:11 & 01:20 & X3.2 & 11748 & N12E76 &~& 02:15 & 06:45 \\
 --    May 15 & 01:25 & 01:48 & 01:58 & X1.2 & 11748 & N11E63 &~& 02:30 & 07:45 \\
  2014 Feb 25 & 00:39 & 00:49 & 01:03 & X4.9 & 11990 & S15E77 &~& 01:10 & 03:10 \\
\tableline
\end{tabular}
\tablenotetext{a}{The locations of the active regions (ARs) are approximate, especially for the ARs on the solar limb.}
\tablenotetext{b}{The approximate time when the first DPFL appears and the last DPFL disappears.}
\end{center}
\end{table}

\begin{table}[htbp]
\begin{center}
\caption{Average width of DPFLs in different \emph{SDO}/AIA EUV channels.} \label{tab:3}
\begin{tabular}{ccccc}
\tableline\tableline
Channel & Char. log (T)\tablenotemark{a} &  Width (km)   &   Number  &  Method\tablenotemark{b}\\
\tableline
AIA 304 &4.7        & 1207 $\pm$ 484 & 3429 &  Auto\\
AIA 171 &5.8        & 1279 $\pm$ 482 & 3747 &  Auto\\
AIA 193 & 6.2, 7.3& 1317 $\pm$ 474 & 3195 &  Auto\\
AIA 211 &6.3        & 1375 $\pm$ 476 & 3684 &  Auto\\
AIA 335 & 6.4       & 1330 $\pm$ 493 & 3263 &  Auto\\
AIA 131 & 5.6, 7.0& 1147 $\pm$ 537 & 2407 &  Auto\\
AIA 171 & 5.8       & 1022 $\pm$ 339 & 4467 &  Manual\\
\tableline
\end{tabular}
\tablenotetext{a}{The characteristic temperatures of AIA EUV channels from \citet{2012SoPh..275...17L}.}
\tablenotetext{b}{The widths are calculated automatically and measured manually, respectively.}
\end{center}
\end{table}

\begin{table}[htbp]
\begin{center}
\caption{Previous results of the temperature (T), electron density (n$_{e}$), half-length (L), cooling time ($\tau$), and ascending velocity (V$_{a}$) of PFLs.} \label{tab:2}
\begin{tabular}{lcccccc}
\tableline\tableline
References & Flare & T    & n$_{e}$     & L     & $\tau$ & V$_{a}$ \\
           & Class & (MK) & (cm$^{-3}$) & (Mm)  & (s) & km~s$^{-1}$ \\
\tableline
 \citet{1995SoPh..156..337S} & X3.9 &  5.5  & $7\times10^{9}$     & 100 & 2000    & 1.4 \\
 \citet{2001SoPh..204...91A} & X5.7 &  27 & $7.9\times10^{11}$  & 27\tablenotemark{a}  & 90, 420\tablenotemark{b} &  -- \\
 \citet{2002SoPh..206..133S} & --\tablenotemark{c} & 1.5   & $5\times10^{8}$     & 30  & 660-780 & 3-8 \\
 \citet{2003SoPh..215..127K} & X9.2 & 9     & $6.9\times10^{10}$  & 29  & 2300    & 1.3 \\
 \citet{2006ApJ...648..712H} & --\tablenotemark{c} & 1.8   & $4\times10^{9}$     & 80  & 900     & 5-7 \\
\tableline
\end{tabular}
\tablenotetext{a}{The value is estimated from a loop radius of 17.5 Mm by the circle assumption.}
\tablenotetext{b}{The dominant times of conductive cooling and radiative cooling respectively.}
\tablenotetext{c}{The flare is not found in the GOES flare list.}
\end{center}
\end{table}

\newpage

\end{document}